\documentclass[preprint,12pt,number,nonatbib]{elsarticle}

\usepackage{amssymb}
\usepackage{graphicx}
\usepackage{amsmath}
\usepackage{amsfonts}
\usepackage{amssymb}
\usepackage{graphicx}
\usepackage{cite}
\usepackage{algorithm}
\usepackage{algorithmic}
\usepackage{array} 
\def\bTet{\mathbf{\Theta}}
\def\cC{\mathbb{C}}

\def\bw{\mathbf{w}}
\def\bH{\mathbf{H}}

\def\bW{\mathbf{W}}

\def\bV{\mathbf{V}}

\def\bh{\mathbf{h}}

\def\bu{\mathbf{u}}
\def\bv{\mathbf{v}}

\allowdisplaybreaks

\journal{Physical Communication}

\begin{document}

\begin{frontmatter}



\title{Joint Design of Transmit Beamforming, IRS Platform, and Power Splitting SWIPT Receivers for Downlink Cellular Multiuser MISO}


\author{Shayan Zargari}
\author{Shahrokh Farahmand\corref{cor1}}
\author{Bahman Abolhassani\fnref{fn1}}

\cortext[cor1]{Corresponding author}
\fntext[fn1]{S. Zargari, S. Farahmand, and B. Abolhassani are with the School
	of Electrical Engineering, Iran University of Science and Technology, Tehran,
	Iran, e-mails: s$\_$zargari@elec.iust.ac.ir, \{shahrokhf,abolhassani\}@iust.ac.ir.}

\begin{abstract}
A multiple antenna base station (BS) with an intelligent reflecting surface (IRS) platform, and several single antenna users are considered in the downlink mode. Simultaneous wireless information and power transfer (SWIPT) is utilized by the BS via transmit beamforming to convey information and power to all devices. Each device applies power splitting (PS) to dedicate separate parts of received power to information decoding and energy harvesting. We formulate a total transmit power minimization problem to jointly design the BS beamforming vectors, IRS phase shifts, and PS ratios at the receivers subject to minimum rate and harvested energy quality of service (QoS) constraints at all the receivers. First, we develop a block coordinate descent algorithm, also known as alternating optimization that can decrease the objective function with every iteration with guaranteed convergence. Afterwards, two low-complexity sub-optimal algorithms that rely on well-known maximum ratio transmission and zero-forcing beamforming techniques are introduced. These algorithms are beneficial when the number of BS antennas and/or number of users are large, or coherence times of channels are small. Simulations corroborate the expectation that by deploying a passive IRS, BS power can be reduced by $10-20$ dBw while maintaining similar guaranteed QoS. Furthermore, even the proposed sub-optimal algorithms outperform the globally optimal SWIPT solution without IRS for a modest number of IRS elements. 
\end{abstract}




\end{frontmatter}


\section{Introduction}
Large growth in the number of wireless devices, as well as their transmission-rates or high-reliability traffic demands, are continually challenging future (five and six-generation) cellular networks to invent more energy-efficient and spectral-efficient solutions and architectures \cite{survey1,survey2,survey3,survey4}. Recently, intelligent reflective surface (IRS) has been introduced as a novel and cost-effective method to enhance the performance of wireless networks in terms of reliability, coverage area, spectral efficiency (SE), and energy efficiency (EE)\cite{2,1}. Compared to the conventional relay schemes, IRS elements are passive and only reflect the incident signals by administering independent phase shifts to the incident wave utilizing low-cost configurable elements, termed as phase shifters \cite{3}. 

Some recent studies have focused on the deployment of the IRS and its optimal design under varying operating conditions. To be specific, \cite{5} investigated sum-rate maximization in an IRS-aided network where the transmitted power of the base station (BS) and reflection coefficients at the IRS were optimized. On the other hand, \cite{7} studied a MISO communication system where the SE was maximized by designing the transmit beamformers and phase shifters at the BS and the IRS. The total transmit power minimization of the IRS network was investigated in \cite{8}, where the transmit beamformers and reflection coefficients were jointly optimized. In addition to the high SE, achieving high EE is a fundamental demand for future wireless generations. As a promising solution, SWIPT technology has been studied comprehensively in the past several years, where each receiver can extract both energy and information simultaneously \cite{12,11,10}. For instance, \cite{12} investigated a downlink (DL) MISO SWIPT-aided system where all users receive information and energy simultaneously by employing power splitting (PS). In particular, the total transmit power was minimized by jointly optimizing beamforming vectors and PS ratios at the BS and receivers, respectively.

To boost system performance, the concept of SWIPT has also been incorporated into the IRS-aided networks, resulting in the notion of IRS-SWIPT. The performance analysis of the IRS-SWIPT with PS receivers was studied in \cite{13} and compared to the conventional decode-and-forward (DF) relaying systems. Specifically, the achievable data rate of the designed system with a single-antenna BS was analyzed, and the corresponding phase shifters values were obtained to maximize it. However, only a single-antenna BS and a single user were considered. In addition, the direct path between the user and BS was assumed non-existent. Two groups of receivers were considered in \cite{14} that either performed energy-harvesting (EH) or information decoding (ID), and the minimum harvested power by EH receivers was maximized. Maximization of the weighted sum of harvested powers for all EH receivers was investigated in \cite{15}. In \cite{16}, transmit power was minimized in the MISO IRS-aided SWIPT system while offering guaranteed minimum QoS in terms of rate and harvested power to the two groups of ID receivers and EH receivers.

Existing contributions, e.g., \cite{14}-\cite{16}, have mainly concentrated on a separate architecture where ID receivers are different from EH ones. In most cases, specifically in internet of things (IoT) applications, same receivers are required to decode data and harvest energy at the same time. While optimization of such combined ID/EH receiver architectures is in high practical demand, they have not been looked at except \cite{13} whose scenario is very narrow in scope. To address this shortcoming, our main contributions can be enumerated as follows:

\begin{enumerate}
	\item[{\bf 1.}] A scenario involving both SWIPT and IRS is investigated where users apply power-splitting (PS) to perform both ID and EH from the same transmissions. It is sought to minimize BS transmit power subject to individual guaranteed QoS in terms of rate and harvested energy. The design involves selecting transmit beamformers, IRS phase shifts, and receivers PS ratios.
	
	\item[{\bf 2.}] Block coordinate descent that is also known as alternating optimization is exploited to derive a solution to the design problem in {\bf 1}. 
	
	\item[{\bf 3.}] While satisfactory in performance, the solution in {\bf 2} suffers from high complexity. For scenarios with large number of BS antennas and/or users or short coherence times, two sub-optimal designs based on maximum ratio transmission (MRT) and zero-forcing (ZF) beamforming are developed. 
	
	\item[{\bf 4.}] Improvements obtained via IRS is investigated numerically. It is revealed that transmission power can be lowered by about $12$ dBw with the incorporation of an IRS with $50$ elements. Furthermore, the proposed sub-optimal designs outperform the globally optimal SWIPT design with no IRS even with a modest number of IRS elements.
\end{enumerate}

The rest of the paper is organized as follows. Section II provides system model and problem formulation. Section III derives the block coordinate descent algorithm. Sub-optimal MRT- and ZF- based solutions are derived in Section IV and numerical results are provided in Section V. Section VI concludes the paper. 

\section{Problem Formulation}
\begin{figure}
	\centering
	\includegraphics[width=0.95\textwidth]{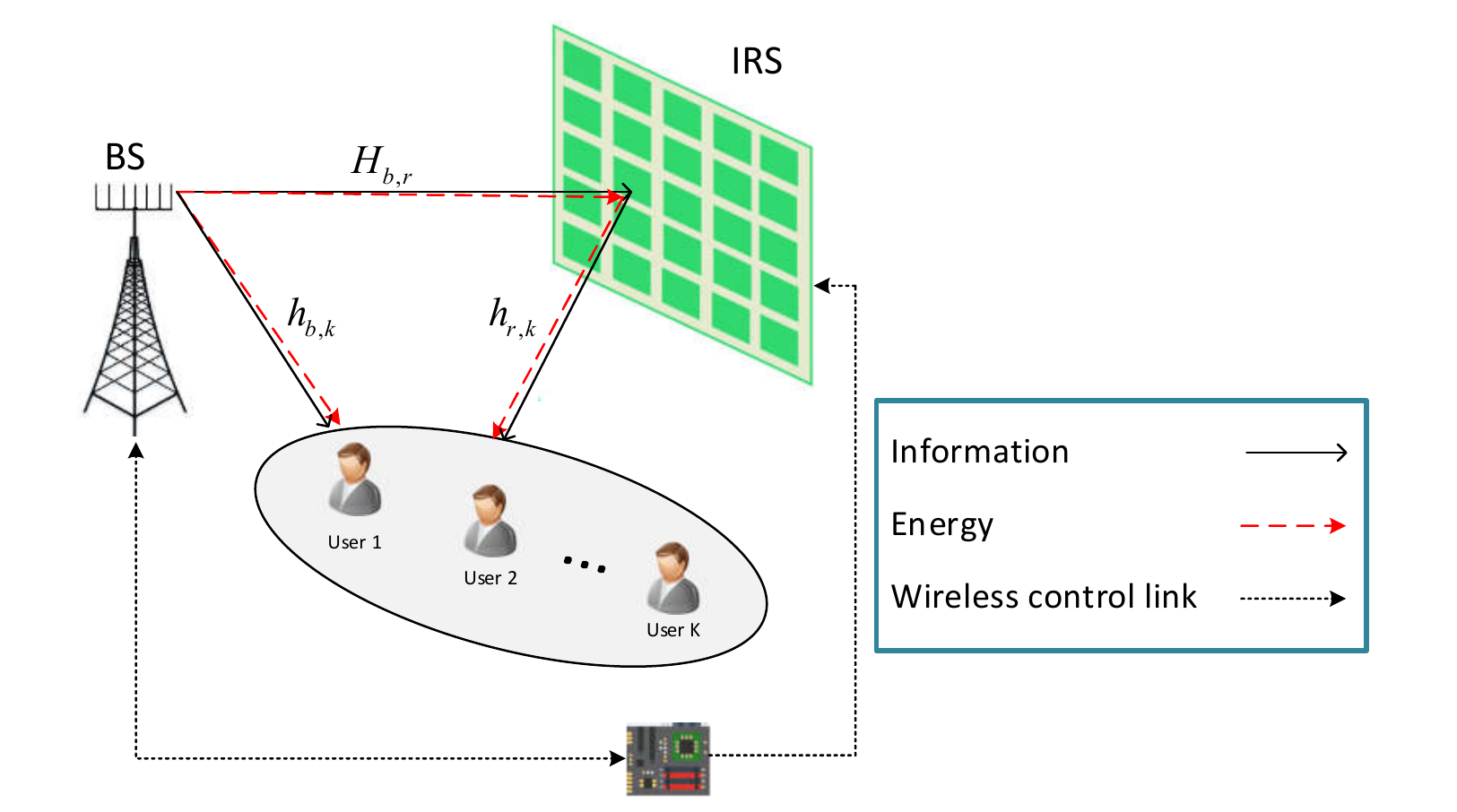}
	\caption{{\small A multi-user MISO IRS-SWIPT communication system.}}
	\label{fig:system_model}
\end{figure}

Downlink transmission from a multi-antenna base station (BS) equipped with $M$ antennas to $K$ single-antenna users is considered, see Fig. \ref{fig:system_model}. In addition to multiple antennas at the BS, an intelligent reflecting surface (IRS) with $N$ elements is exploited to further improve communications. For all the channels between BS, IRS, and users, a block fading model is considered, which means that the channel remains constant within a coherence time and changes independently afterwards. Furthermore, full channel state information (CSI) is assumed at the BS. This can be achieved provided that coherence time is large enough so that all channel gains can be estimated and fed back to the BS. To enable CSI acquisition at the IRS, an IRS controller is employed. Orthogonal frequency-division multiplexing (OFDM) is used to transmit signals to each of the corresponding $K$ users, and one such frequency resource is utilized by all users. The transmitted complex baseband signal at the BS can be expressed by 
\begin{equation}\label{transmit signal}
\mathbf{x} =\sum\limits_{k = 1}^K {{s_k}{\mathbf{w}_k}}, \\
\end{equation}
where $s_{k}\in\cC$ is the information-bearing symbol for user $k$, which is a zero-mean unit variance random variable and independent across users. Furthermore, ${\mathbf{w}}_{k}\in \mathbb{C}^{M \times 1}$ represents the transmit beamforming vector corresponding to user $k$. 

In addition to CSI estimation and feedback, IRS controller adjusts the reflection coefficients of IRS elements at the instruction of BS. IRS is assumed passive, hence its reflection coefficient amplitudes are all equal one and they only perform phase shifts. The reflection coefficient matrix at the IRS is donated by ${\mathbf{\Theta}}=\text{diag} (e^{j\theta_{1}}, e^{j\theta_{2}},\ldots, e^{j\theta_{N}})$, where $\theta_n\in(0,2\pi]$ represents the phase shift of $n$-th unit, or element. The complex conjugate of channel gains between BS and user $k$ is denoted by ${\mathbf{h}}_{b,k}\in \mathbb{C}^{M\times 1}$. Similarly, the complex conjugate of channel gains between the IRS and user $k$ is represented by ${\mathbf{h}}_{r,k}\in \mathbb{C}^{N\times 1}$. The complex channel gains between the BS and the IRS are given by ${\mathbf{H}}_{b,r}\in \mathbb{C}^{N \times M}$. The received signal of user $k$ can be then written as
\begin{equation}\label{received}
{y_k} = {\mathbf{h}}_k^H\left(\sum\limits_{k = 1}^K {{s_k}{\mathbf{w}_k}}\right) + {n_k},
\end{equation}
where $\mathbf{h}_k^H=\mathbf{h}_{b,k}^H+\mathbf{h}_{r,k}^H\mathbf{\Theta}\mathbf{H}_{b,r}$ is the equivalent channel gain from the BS to user $k$, and $(.)^H$ denotes the Hermitian transpose operation. While the first term represents the direct link between the BS and user $k$, second term specifies the reflected path by the IRS. Furthermore, $n_{k}\sim \mathcal{C}\mathcal{N}(0,\,\sigma_{k}^2)$ denotes the antenna noise at the user $k$ which is assumed to be circularly symmetric complex Gaussian (CSCG).
The BS strives to provide all $K$ users with energy as well as information leading to a SWIPT scenario. Among various SWIPT architectures, a power-splitting (PS) one is assumed. Subsequently, each device uses an adjustable power splitter to divide received signal into two streams. A power ratio of $(1-\rho_k)$ is dedicated to energy harvesting (EH), while the remaining $\rho_k$ portion is exploited for information decoding (ID). Consequently, the input signal at the ID section can be stated as
\begin{equation}\label{ID}
y_{k}^{\text{ID}}=\sqrt{\rho_{k}}\left({\mathbf{h}}_k^H\sum\limits_{j = 1}^K {{\mathbf{w}_k}{s_k}} + {n_k}\right)+z_{k},
\end{equation}
where $\rho_k \in [0,1]$ denotes the PS ratio at user $k$, and $z_{k}\sim \mathcal{C}\mathcal{N}(0,\,\delta_{k}^2)$ is the baseband signal processing noise introduced at the ID section which is CSCG. User $k$ decodes its information treating other users as noise. Therefore, the received SINR at the $k$-th user can be expressed as
\begin{equation}\label{sinr}
\text{SINR}_k= \frac{{{\rho _k}{{\left| {\mathbf{{\rm{ {\mathbf{h}}}}}}_{k}^H {\mathbf{{{w}}}}_k \right|}^2}}}{{{\rho _k}\sum\limits_{\scriptstyle \atop\scriptstyle i \ne k} {{{\left|  {\mathbf{{\rm{ {\mathbf{h}}}}}}_{k}^H {\mathbf{{{w}}}}_i \right|}^2} + {\rho _k}\sigma _k^2 + \delta _k^2} }}.\\
\end{equation}

In addition, the received signal at the EH module is given by
\begin{equation}\label{EH}
y_{k}^{\text{EH}}=\sqrt{1-\rho_{k}}\left( {\mathbf{{\rm{ {\mathbf{h}}}}}}_{k}^H{\mathbf{x}} + {n_k}\right).\\
\end{equation}

Moreover, we assume that the amount of harvested power at the EH section is linearly proportional to the received power and can be written as
\begin{equation}\label{harvested power}
{P_k}=\eta_k(1-\rho_{k})\left(\sum\limits_{i = 1}^K {{\left|  {\mathbf{{\rm{ {\mathbf{h}}}}}}_{k}^H {\mathbf{{{w}}}}_i \right|}^2}\right),
\end{equation}
where $\eta_k \in (0,1]$ is the energy conversion efficiency for user $k$. Since the noise power is negligible, so it is not considered in (6). 

Our goal is to jointly design beamforming vectors $\{\bw_k\}_{k=1}^K$, IRS phase shifts $\{\theta_n\}_{n=1}^{N}$, and users power-splitting ratios $\{\rho_k\}_{k=1}^K$ so that the total transmitted power at the BS is minimized while maintaining minimum rate and energy harvesting constraints for all $K$ users. The corresponding optimization problem can be mathematically formulated as follows:
\begin{subequations}
	\begin{align}
	\text{P1}: & \underset{\rho_k,{\mathbf{w}}_k,{\mathbf{\Theta}}} {\text{min}} \quad \sum_{k=1}^{K}{\left\| {\mathbf{w}}_k \right\|^2}\\
	&\text{s.t.} \quad \frac{{{\rho _k}{{\left| {{\mathbf{h}}_{k}^\text{H}{{\mathbf{w}}_k}}  \right|}^2}}}{{{\rho _k}\sum\limits_{\scriptstyle \atop\scriptstyle i \ne k} {{{\left|  {{\mathbf{h}}_{k}^\text{H}{{\mathbf{w}}_i}}  \right|}^2} + {\rho _k}\sigma _k^2 + \delta _k^2} }}\geq \gamma_{k},\: \forall k,\label{p1-1}\\
	&\quad\quad \eta_{k}(1-\rho_{k})(\sum\limits_{i = 1}^K {{{\left| {{\mathbf{h}}_{k}^\text{H}{{\mathbf{w}}_i}}  \right|}^2}})\geq e_{k}, \: \forall k,\label{p1-2}\\
	&\quad\quad 0 < \rho_{k} < 1, \: \forall k,\label{p1-3}\\
	&\quad\quad 0 < {\theta _n} \le 2\pi, \: \forall n,\label{p1-4}
	\end{align}
\end{subequations}
where $\gamma_{k}$ and $\text{e}_{k}$ are given minimum targets for SINR and harvested power at each receiver, respectively. Constraints (\ref{p1-1}) and (\ref{p1-2}) denote the minimum SINR and EH at each user. The given minimum SINR constraint in \eqref{p1-1} is equivalent to a minimum rate constraint if we add one to both sides and take logarithms afterwards. Constraints (\ref{p1-3}) and (\ref{p1-4}) are inherent limits for PS ratios and phase shifters. It should be noted that positive SINR and harvested power thresholds (i.e., $\gamma_{k}>0$, and $\text{e}_{k}>0$) are considered. Hence, $\rho_k$ needs to satisfy strict inequality $0<\rho_k<1$ expressed by constraint (\ref{p1-3}).

\section{An Iterative Solution via Block Coordinate Descent}
Before proceeding to solve P1, we offer some background on the difficulty and available solutions for its simpler variants. When no IRS is present and SWIPT is not utilized, P1 becomes a simple beamforming design problem that minimizes total BS transmit power subject to rate constraints for all users. This problem has the following form
\begin{subequations}
	\begin{align}
	\text{Q1}: & \underset{{\mathbf{w}}_k} {\text{min}} \quad \sum_{k=1}^{K}{\left\| {\mathbf{w}}_k \right\|^2}\\
	&\text{s.t.} \quad \frac{{{{\left| {{\mathbf{h}}_{k}^\text{H}{{\mathbf{w}}_k}}  \right|}^2}}}{{\sum\limits_{\scriptstyle \atop\scriptstyle i \ne k} {{{\left|  {{\mathbf{h}}_{k}^\text{H}{{\mathbf{w}}_i}}  \right|}^2} + \sigma _k^2 + \delta _k^2} }}\geq \gamma_{k},\: \forall k. \label{q1-1}
	\end{align}
\end{subequations}
Here $\bh_k$s are known and fixed, unlike when IRS is present. Multiplying both sides of \eqref{q1-1} by its denominator, Q1 becomes a quadratically-constrained quadratic program (QCQP). QCQPs can range from simple convex problems to very difficult NP-hard ones. While Q1 is not convex, it is not NP-hard and efficient solutions to reach its global optimum exist. Specifically, \cite{original_beamforming} provides an iterative algorithm, which alternates between optimizing beamformers directions and their powers. Closed-form solutions are provided for every step and convergence to the global optimum is proved. Furthermore, \cite{original_beamforming} proves that a semi-definite relaxation (SDR) of Q1 is tight and always provides a rank one solution even when the rank constraint is dropped. This is in contrast to the loosely speaking dual problem (not mathematically defined dual) of designing beamformers that maximize sum-rate subject to individual power constraints, which has been proven to be NP-hard \cite{np-hard}.

A PS SWIPT scenario was added to Q1 by \cite{12} leading to the following problem
\begin{subequations}
	\begin{align}
	\text{Q2}: & \underset{\rho_k,{\mathbf{w}}_k} {\text{min}} \quad \sum_{k=1}^{K}{\left\| {\mathbf{w}}_k \right\|^2}\\
	&\text{s.t.} \quad \frac{{{\rho _k}{{\left| {{\mathbf{h}}_{k}^\text{H}{{\mathbf{w}}_k}}  \right|}^2}}}{{{\rho _k}\sum\limits_{\scriptstyle \atop\scriptstyle i \ne k} {{{\left|  {{\mathbf{h}}_{k}^\text{H}{{\mathbf{w}}_i}}  \right|}^2} + {\rho _k}\sigma _k^2 + \delta _k^2} }}\geq \gamma_{k},\: \forall k\\
	&\quad\quad \eta_{k}(1-\rho_{k})(\sum\limits_{i = 1}^K {{{\left| {{\mathbf{h}}_{k}^\text{H}{{\mathbf{w}}_i}}  \right|}^2}})\geq e_{k}, \: \forall k\\
	&\quad\quad 0 < \rho_{k} < 1, \: \forall k,
	\end{align}
\end{subequations}
Although the constraints in Q2 are no longer quadratic, Q2 has curiously maintained the specific structure of Q1. Subsequently, \cite{12} proved that a SDR of Q2 is also tight and provides a rank one optimum solution. Furthermore, the SDR is convex and can be solved efficiently. 

Unlike Q1 and Q2, P1 can not be easily relaxed into a convex problem via SDR as the IRS phase shifts enter the problem through $\bh_k$'s, which appear quadratically alongside quadratic $\bw_k$s. In order to tackle this challenge, P1 is solved via block coordinate descent (BCD), also known as alternating optimization (AO), which is proved to reduce the objective function at every iteration with guaranteed convergence. In the first stage, IRS phase shifts are fixed while the beamformers and PS ratios are optimized via SDR. In the second stage, beamformers and PS ratios are fixed and IRS phase shifts are optimized.\\
\subsection{Optimizing $\mathbf{w}_k$ and $\rho_k$ with given $\mathbf{\Theta}$} 
With $\mathbf{\Theta}$ fixed, $\bh_k$'s become fixed and P1 reduces to that of beamforming with PS SWIPT in Q2. Upon defining $\bW_k=\bw_k\bw_k^H$, this problem can be relaxed as
\begin{subequations}
	\begin{align}
	\text{P2}:&  \underset{\rho_k,{\mathbf{W}}_k} {\text{min}} \quad \sum_{k=1}^{K} \text{Tr}({\mathbf{W}}_k)\\
	&\text {s.t.} \quad \frac{{\text{Tr}({{\mathbf{H}}_k}{{\mathbf{W}}_k})}}{{{\gamma _k}}} - \sum\limits_{\scriptstyle \atop\scriptstyle i \ne k} {\text{Tr}({{\mathbf{H}}_k}{{\mathbf{W}}_i})}  \ge \sigma _k^2 + \frac{{\delta _k^2}}{{{\rho _k}}},\: \forall k,\label{p2-1}\\
	&\quad\quad \sum\limits_{i = 1}^K {{ \text{Tr}({{\mathbf{H}}_k}{{\mathbf{W}}_i}) } \ge \frac{{{e_k}}}{{{\eta _k}(1 - {\rho _k})}} }, \: \forall k,\label{p2-2}\\
	&\quad\quad {\mathbf{W}}_k\geq 0,\: \forall k,\label{p2-3}\\
	&\quad\quad 0 < \rho_{k} < 1, \: \forall k.\label{p2-4}
	\end{align}
\end{subequations}

It is notable that the right-hand side of constraints in (\ref{p2-1}) and (\ref{p2-2}) are both convex functions given the fact that both $\frac{1}{\rho_{k}}$ and $\frac{1}{1-\rho_{k}}$ are convex and the left-hand side of these constraints are affine. As seen, rank one constraint is dropped due to relaxation. If $\mathbf{W}_k^\ast$ which is the optimum solution to P2 satisfies rank one constraint, the optimal beamforming weight $\bw_k$ can be found by an eigenvalue decomposition (EVD) of $\mathbf{W}_k^\ast$. For a fixed and given $\mathbf{\Theta}$, we know from \cite{12} that $\mathbf{W}_k^\ast$ is guaranteed to have rank one. Subsequently, EVD of $\mathbf{W}_k^\ast$ leads to the optimal solution. $\bW_k^{\ast}$ itself can be obtained by employing an interior-point algorithm \cite{17}, and using computer software tool, e.g., CVX \cite{18}. 

\subsection{Optimizing $\mathbf{\Theta}$ with given $\mathbf{w}_k$ and $\rho_k$} 
With beamforming weights $\bw_k$ fixed, the objective function in P1 becomes fixed, so the problem is changed into a feasibility check for $\bTet$. Let us define, ${\mathbf{u}}=[e^{j\theta_{1}},e^{j\theta_{2}},...e^{j\theta_{N}}]^\text{H}$ and apply the change of variables ${\mathbf{h}}_{b,k}^\text{H}{\mathbf{w}}_i^ * = {a_{b,i}}$ and ${\mathbf{h}}_{r,k}^\text{H}{\mathbf{\Theta}} {\mathbf{H}}_{b,r}{\mathbf{w}}_i^ *  = {{\mathbf{u}}^\text{H}}{{\mathbf{b}}_i}$, where ${{\mathbf{b}}_i} = \text{diag}({\mathbf{h}}_{r,k}^\text{H}){\mathbf{H}_{b,r}}{\mathbf{w}}_i^ * $. Then P1 can be rewritten as follows
\begin{subequations}
	\begin{align}
	\text{P3:} ~~&\text{Find}\quad  \mathbf{u}\\
	&\text{s.t.} \quad \frac{{{{\left| {{{\mathbf{u}}^\text{H}}{{\mathbf{b}}_k} + {a_{b,k}}} \right|}^2}}}{{\sum\limits_{\scriptstyle \atop\scriptstyle i \ne k} {{{\left| {{{{\mathbf{u}}^\text{H}}}{{\mathbf{b}}_i} + {a_{b,i}}} \right|}^2} + \sigma _k^2 + \frac{{\delta _k^2}}{{\rho _k^ * }}} }}\geq \gamma_{k},\: \forall k,\label{p3-1}\\
	&\quad\quad \eta_{k}(1-\rho_{k}^*)\big(\sum\limits_{i = 1}^K  {{{\left| {{{\mathbf{u}}^\text{H}}{{\mathbf{b}}_i} + {a_{b,i}}} \right|}^2}}\big)\geq e_{k}, \: \forall k,\label{p3-2}\\
	&\quad\quad |{{u}}_{n}|=1,\: \forall n.\label{p3-3}
	\end{align}
\end{subequations}
P3 maintains quadratic inequality and equality constraints. Subsequently, we can use SDR to relax P3. Specifically, we introduce ${\mathbf{v}}=[{\mathbf{u}};{\mathbf{1}}]$ and ${{\mathbf{G}}_i} = [{{\mathbf{b}}_i}{\mathbf{b}}_i^H,{{\mathbf{b}}_i}a_{b,i}^{H};{a_{b,i}}{\mathbf{b}}_i^H,0]$. Furthermore, we define ${{\mathbf{V}}}={{\mathbf{v}}}{\mathit{{\mathbf{v}}}}^H$ which requires to satisfy ${\mathbf{V}}\geq\mathbf{0}$ and $\text{Rank}({\mathbf{{V}}})=1$. Dropping rank one constraint, (P3) is relaxed into
\begin{subequations}
	\begin{align}
	\text{P4:}~~&\text{Find}  \quad \mathbf{V}\\
	&\text{s.t.} \quad \frac{{\text{Tr}({{\mathbf{G}}_k}{\mathbf{V}})}}{{{\gamma _k}}} - \sum\limits_{\scriptstyle i = 1\atop\scriptstyle i \ne k}^K {\text{Tr}({{\mathbf{G}}_i}{\mathbf{V}})}  \ge \sigma _k^2 + \frac{{\delta _k^2}}{{\rho _k^ * }},\: \forall k,\\
	&\quad\quad \sum\limits_{i = 1}^K {\text{Tr}({{\mathbf{G}}_i}{\mathbf{V}})}  \ge \frac{{{e_k}}}{{{\eta _k}(1 - {\rho _k})}} , \: \forall k,\\
	&\quad\quad {\mathbf{V}}_{n,n}=1,~ n=1,...,N+1, \quad  \mathbf{{{V}}} \geq 0.
	\end{align}
\end{subequations}

P4 is a standard SDP and can be solved with optimization solvers such as CVX \cite{18}. While we may be tempted to iteratively solve P2 and P4, the iterations can get stuck even in the first round because the same initialization $\bTet^{(0)}$ used in P2 remains feasible in P4. Hence, a mechanism should be developed to ensure $\bTet$ keeps varying in P4. To accommodate such mechanism, we consider the following optimization problem instead of (P4):  
\begin{subequations}
	\begin{align}
	\text{P5}: &  \underset{\bV,\alpha_k,\phi_k} {\text{max}} \quad \sum_{k=1}^{K} \alpha_k+\lambda \phi_k\\
	&\text{s.t.} \quad \frac{{\text{Tr}({{\mathbf{G}}_k}{\mathbf{V}})}}{{{\gamma _k}}} - \sum\limits_{\scriptstyle i = 1\atop\scriptstyle i \ne k}^K {\text{Tr}({{\mathbf{G}}_i}{\mathbf{V}})}  \ge \sigma _k^2 + \frac{{\delta _k^2}}{{\rho _k^ * }}+\alpha_k,\: \forall k,\label{p5-1}\\
	&\quad\quad \sum\limits_{i = 1}^K {\text{Tr}({{\mathbf{G}}_i}{\mathbf{V}})}  \ge \frac{{{e_k}}}{{{\eta _k}(1 - {\rho _k})}} +\phi_k, \: \forall k,\label{p5-2}\\
	&\quad\quad {\mathbf{V}}_{n,n}=1,~ n=1,...,N+1, \quad  {{\mathbf{V}}} \geq 0, \\
	& \quad \quad \alpha_k \geq 0,~\phi_k\geq 0,\quad \forall k=1,\ldots,K.
	\end{align}
\end{subequations}

P5 is designed to select $\bTet$ (through $\bV$) so as to maximize the SINR and EH margins from the minimum required values specified in P2. Hence, it looks to somehow optimize $\bTet$ instead of just providing a feasible point. It should be noted that the optimal solution of P5 is a feasible solution of P4. 

It is notable that the SDR relaxation appearing in P5 may not be tight. Therefore, 
optimal $\bV$ may not be rank one. While Gaussian randomization can be employed to obtain a feasible solution in the form of $\bv=[\bu; 1]$, we utilize a simple alternative. Indeed, we perform eigenvalue decomposition (EVD) of $\bV$ and select the eigenvector corresponding to the largest eigenvalue as $\bv$. Next, we check if this $\bv=[\bu; 1]$ satisfies the constraints in P3. If it does, then we are done. If it does not, then we move on to the next largest eigenvalue and continue until a feasible solution is found.  In the end, P2 and P5 can be solved iteratively instead of the main problem P1. The block coordinate descent algorithm to solve P1 is summarized in Algorithm 1. The following proposition offers its performance guarantee.

\begin{algorithm}[t]
	\caption{Block Coordinate Descent for P1}
	\begin{algorithmic}[1]
		\renewcommand{\algorithmicrequire}{\textbf{Input:}}
		\renewcommand{\algorithmicensure}{\textbf{Output:}}
		\REQUIRE Set iteration number $i=0$ and initialize the phase shifts to $\mathbf{\Theta}^{(0)}$.\\    
		\STATE \textbf{While} $\bTet^{(i+1)}\neq\bTet^{(i)}$\:\textbf{do}\\
		\STATE \quad Solve P2 for given $\mathbf{\Theta}^{(i)}$, and obtain the optimal \\ \quad solutions as $\{\mathbf{w}_k^{(i+1)},\rho_k^{(i+1)}\}$.
		\STATE \quad Solve P5 for given $\{\mathbf{w}_k^{(i+1)},\rho_k^{(i+1)}\}$, and  denote the \\ \quad feasible  solution after applying EVD by  $\mathbf{\Theta}^{(i+1)}$.
		\STATE \quad set $i\leftarrow i+1$;
		\STATE  \textbf{end while}
		\ENSURE return solutions $\{\mathbf{w}_k^\ast,\rho_k^\ast,\mathbf{\Theta}^\ast\}$.
	\end{algorithmic}
\end{algorithm}

\noindent {\textbf{Proposition 1}}:
Algorithm 1 iterations yield a non-increasing sequence of objective values with guaranteed convergence.

\textit{Proof}: Please check Appendix A.

\section{Low-complexity Sub-optimal Solutions}
While Algorithm 1 provides a satisfactory performance, its complexity might be too demanding for scenarios involving large number of antennas and/or users, or short coherence times. To address this limitation, we resort to sub-optimal low-complexity alternatives. Specifically, we utilize simple beamforming methods such as maximum ratio transmission (MRT) or zero-forcing (ZF) and optimize only the beamforming powers and PS ratios. This reduces the parameters dimension to $2K$ compared to that of P2 which has $(M^2+1)K$ parameters. We focus on MRT first and investigate ZF afterwards.

\subsection{MRT Beamformer}
MRT beamformer is given by
\begin{equation}
\mathbf{w}^{(\text{MRT})}_k=\frac{\sqrt{P_k}\mathbf{h}_k}{\|\mathbf{h}_k\| }=\sqrt{P_k}\bar{\mathbf{w}}_k.\label{mrt}
\end{equation}

Using $\bw_k$ as defined in (\ref{mrt}) in Q2 yields
\begin{subequations}
	\begin{align}
	\text{P6}: &  \underset{\rho_k,P_k} {\text{min}} \quad \sum_{k=1}^{K}{P_k}\\
	&\text{s.t.} \quad (1+\gamma_{k}){ P_k\left| \mathbf{h}_{k}^H\bar{\bw}_k  \right|^2}\geq \label{14b}  \\ & \qquad \qquad \frac{\gamma_k}{\rho_{k}}\delta_{k}^2+\gamma_k\left(\sum\limits_{i = 1}^K P_i{{{\left| {{\mathbf{h}}_{k}^\text{H}{\bar{{\mathbf{w}}}_i}}  \right|}^2}}+\sigma^2_{k}\right),\: \forall k, \nonumber\\
	&\quad\quad \eta_{k}(1-\rho_{k})\left(\sum\limits_{i = 1}^K P_i{{{\left| {{\mathbf{h}}_{k}^\text{H}{\mathbb{{\bar{\mathbf{w}}}}_i}}  \right|}^2}}\right)\geq e_{k}, \: \forall k,\\
	&\quad\quad P_k\geq 0,\: \forall k,\label{14d}
	\quad \quad 0 < \rho_{k} < 1, \: \forall k.
	\end{align}
\end{subequations}
With $\bTet$ and thus $\bh_k,\bar{\bw}_k$s fixed, P6 is convex. Indeed, P6 can be transformed into a second-order cone program (SOCP), which can be efficiently solved \cite{19}. We show how to reformulate P6 into a SOCP next. One of the constraint forms which can be represented by SOCs is the restricted hyperbolic constraints. They have the form of $\mathbf{ x}^H\mathbf{ x}\leq yz$, which are equivalent to the following form
\begin{equation}
\left\| {\begin{array}{*{20}{c}}{2\mathbf{x}}\\{y - z}\end{array}} \right\| \le y + z,
\end{equation}
where $\mathbf{x}\in \mathbb{C}^{M\times1}$ and $y,z\geq 0$ are real non-negative scalars. Let us define $\bar{P}_k:=\sum_{i = 1}^K P_i{{{\left| {{\mathbf{h}}_{k}^\text{H}{\bar{{\mathbf{w}}}_i}}  \right|}^2}}$. Next, we  introduce $z_k:=(1+\gamma_{k}){ P_k{{{\left| {{\mathbf{h}}_{k}^\text{H}{\bar{{\mathbf{w}}}_k}}  \right|}^2}}}-\gamma_k(\bar{P}_k+\sigma_k^2)$, where the slack variable $z_k\geq 0$ because $\frac{\gamma_k}{\rho_{k}}\delta_{k}^2\geq 0$. Otherwise, (\ref{14b}) becomes infeasible. Then, we replace $z_k$ in (\ref{14b}) which yields $\rho_{k}z_k\geq \gamma_{k}\delta_{k}^2$. Consequently, the SOCP form can be formulated as
\begin{subequations}
	\begin{align}
	\text{P8}: &  \underset{\rho_k,P_k,\bar{P}_k,z_k} {\text{min}} \quad \sum_{k=1}^{K}{P_k}\\
	&\text{s.t.} \quad (1+\gamma_{k}){ P_k{{{\left| {{\mathbf{h}}_{k}^\text{H}{\bar{{\mathbf{w}}}_k}}  \right|}^2}}}= z_k+\gamma_k(\bar{P}_k+\sigma_k^2),\: \forall k,\label{16b}\\
	&\quad\quad 	\left\| {\begin{array}{*{20}{c}}{{2}\sqrt{\gamma_{k}\delta^2_{k}}}\\{z_k- \rho_{k}}\end{array}} \right\| \le z_k + \rho_{k},\: \forall k,\\
	&\quad\quad 	\left\| {\begin{array}{*{20}{c}}{{2}\sqrt{\frac{e_k}{\eta_{k}}}}\\{(1-\rho_{k})-\bar{P}_k}\end{array}} \right\| \le(1-\rho_{k})+\bar{P}_k,\: \forall k,\\
	&\quad\quad z_k\geq 0,~P_k\geq 0,~ 0 \leq \rho_k \leq 1,\: \forall k,\\
	& \quad \quad \bar{P}_k:=\sum_{i = 1}^K P_i{{{\left| {{\mathbf{h}}_{k}^\text{H}{\bar{{\mathbf{w}}}_i}}  \right|}^2}},~\forall k.
	\end{align}
\end{subequations}
P8 is a SOCP, which can be efficiently solved. Therefore, our first sub-optimal algorithm iteratively solves P8 and P5 where beamformers' directions are selected via MRT. MRT-based sub-optimal algorithm is summarized in Algorithm 2. 	

\begin{algorithm}[t]
	\caption{BCD with MRT beamforming}
	\begin{algorithmic}[1]
		\renewcommand{\algorithmicrequire}{\textbf{Input:}}
		\renewcommand{\algorithmicensure}{\textbf{Output:}}
		\REQUIRE Set iteration number $i=0$ and initialize the phase shifts to $\mathbf{\Theta}^{(0)}$.\\    
		\STATE \textbf{While} $\bTet^{(i+1)}\neq\bTet^{(i)}$\:\textbf{do}\\
		\STATE \quad Solve P8 for given $\mathbf{\Theta}^{(i)}$, and obtain the optimal \\ \quad solutions as $\{\rho_k^{(i+1)},P_k^{(i+1)}\}$.
		\STATE \quad For given $\{\rho_k^{(i+1)},P_k^{(i+1)}\}$, evaluate $\bw_k^{(\text{MRT})}$ as in \eqref{mrt} \\ \quad then denote the feasible  solution of P5 after \\ \quad applying EVD by $\mathbf{\Theta}^{(i+1)}$.
		\STATE \quad set $i\leftarrow i+1$;
		\STATE  \textbf{end while}
		\ENSURE return solutions $\{P_k^{*},\rho_k^\ast,\mathbf{\Theta}^\ast\}$.
	\end{algorithmic}
\end{algorithm}

\subsection{ZF Algorithm}
ZF-based beamforming can eliminate interference when $M\geq K$. As a result, P1 reduces to the following problem
\begin{subequations}
	\begin{align}
	\text{P9}: &  \underset{\rho_k,{\mathbf{w}}_k} {\text{min}} \quad \sum_{k=1}^{K}{\left\| {\mathbf{w}}_k \right\|^2}\\
	&\text{s.t.} \quad \frac{{{\rho _k}{{\left| {{\mathbf{h}}_{k}^\text{H}{{\mathbf{w}}_k}}  \right|}^2}}}{{ {  {\rho _k}\sigma _k^2 + \delta _k^2} }}\geq \gamma_{k},\: \forall k,\label{p9-1}\\
	&\quad\quad \eta_{k}(1-\rho_{k}){{{\left| {{\mathbf{h}}_{k}^\text{H}{{\mathbf{w}}_k}}  \right|}^2}}\geq e_{k}, \: \forall k,\label{p9-2}\\
	&\quad\quad \bar{\mathbf{H}}^H_k\mathbf{w}_k=0,\: \forall k,\\
	&\quad\quad 0 < \rho_{k} < 1, \: \forall k,
	\end{align}
\end{subequations}
where $\bar{\mathbf{H}}_k \buildrel \Delta \over = [{\mathbf{ h}_1},...,{\mathbf{h}_{k - 1}},{\mathbf{h}_{k + 1}},...,{\mathbf{ h}_K}] \in {\mathbb{C}^{M \times (K - 1)}}$. The optimal solution for (P9) has a closed-form as determined by \cite{12}. We explain the main ideas for finding the closed-form solution here and leave the derivations to \cite{12}.

First, constraints in P9 are decoupled across users and hence beamformer for every user can be optimized separately. Second, both constraints \eqref{p9-1} and \eqref{p9-2} should hold with equality at the optimum for any given $k$. Otherwise, we can change $\rho_k$ a little bit so that both constraints are held with inequality then decrease $\|\bw_k\|^2$ while ensuring all constraints are satisfied. As a result, for any given $k$, \eqref{p9-1} and \eqref{p9-2} become two equations in two unknowns which are $\rho_k$ and $|\bh_k^H\bw_k|^2$. Omitting $|\bh_k^H\bw_k|^2$ from one and using into the second introduces a quadratic equation for $\rho_k$ which has one positive solution. After finding optimal $\rho_k$, optimal $\bw_k$ is found by projecting $\bh_k$ into the null space of $\bar{\bH}_k$ followed by adjusting its $\ell_2$ norm to ensure it equals the optimal value of $|\bh_k^H\bw_k|^2$ found via \eqref{p9-1} and \eqref{p9-2}. Closed-form solution for P9 is summarized in Algorithm 3, where $\mathbf{U}_k$ indicates the orthogonal basis of the null space of $\bar{\mathbf{H}}^H_k$. Finally, Algorithm 4 summarizes the sub-optimal ZF-based block coordinate descent solution to P1.

\begin{algorithm}[t]
	\caption{Closed-Form Solution for P9}
	\begin{algorithmic}[1]
		\renewcommand{\algorithmicrequire}{\textbf{Input:}}
		\renewcommand{\algorithmicensure}{\textbf{Output:}}    
		\STATE For given $\bh_k$s, obtain the optimal solutions $\{\mathbf{w}_k^\ast,\rho_k^\ast\}$ as follows \cite{12}:
		\STATE \quad Set $\mathbf{U}_k=$null$(\bar{\mathbf{H}}^H_k)$, $\forall k$.
		\STATE \quad Set $\kappa _k = \frac{{{e_k}}}{{{\eta _k}({\gamma _k} + 1)\sigma _k^2}}$ and $\vartheta _k= \frac{{{\gamma _k}\delta _k^2}}{{({\gamma _k} + 1)\sigma _k^2}}$, $\forall k$.
		\STATE \quad Set $\rho _k^* = \frac{{ - ({\kappa _k} + {\vartheta _k} - 1) + \sqrt {{{({\kappa _k} + {\vartheta _k} - 1)}^2} + 4{\vartheta _k}} }}{2}$, $\forall k$.
		\STATE \quad Set $\bw_k^* = \sqrt {{\gamma _k}(\sigma _k^2 + \frac{{\delta _k^2}}{{\rho _k^*}})} \frac{{{\mathbf{ U}_k}\mathbf{U}_k^H{\mathbf{h}_k}}}{{\mathbf{ h}_k^H{\mathbf{ U}_k}\mathbf{ U}_k^H{\mathbf{h}_k}}}$, $\forall k$.
	\end{algorithmic}
\end{algorithm}
\begin{algorithm}[t]
	\caption{BCD with ZF beamforming}
	\begin{algorithmic}[1]
		\renewcommand{\algorithmicrequire}{\textbf{Input:}}
		\renewcommand{\algorithmicensure}{\textbf{Output:}}
		\REQUIRE Set iteration number $i=0$ and initialize the phase shifts to $\mathbf{\Theta}^{(0)}$.\\    
		\STATE \textbf{While} $\bTet^{(i+1)}\neq\bTet^{(i)}$\:\textbf{do}\\
		\STATE \quad Solve Algorithm 3 for given $\mathbf{\Theta}^{(i)}$, and obtain the optimal solutions as $\{\rho_k^{(i+1)},\bw_k^{(i+1)}\}$.
		\STATE \quad Solve (P5) for given $\{\rho_k^{(i+1)},\bw^{(i+1)}_k\}$, and  denote the \\ \quad feasible  solution after applying EVD by  $\mathbf{\Theta}^{(i+1)}$.
		\STATE \quad set $i\leftarrow i+1$;
		\STATE  \textbf{end while}
		\ENSURE return solutions $\{\bw^{*}_k,\rho_k^\ast,\mathbf{\Theta}^\ast\}$.
	\end{algorithmic}
\end{algorithm}

\begin{table}[t]
	\renewcommand{\arraystretch}{1.05}
	\centering
	\caption{Simulation Parameters}
	\label{table-notations}
	\begin{tabular}{| c| c| }    
		%\begin{tabular}{| c| l| }    
		\hline
		\textbf{Parameters}& \textbf{Values}\\\hline        
		%%%%%%%%%%%parameters%%%%%%%%%%%%%%%%%%
		BS location & (0m,0m)  \\ \hline
		IRS location & (10m,0m)  \\ \hline
		Rectangular Dimensions & 5m$\times$5m \\ \hline
		Path-loss for the direct link& -30-20*log10(d) \\ \hline
		Path-loss for the reflected link& -30-20*log10(d) \\ \hline
		energy conversion efficiency & $\eta_k=\eta=0.5$  \\ \hline
		the antenna noise & $\sigma_k^2=\sigma^2=-70~\text{dBm}$  \\ \hline
		additional noise at the ID & $\delta_{k}^2=\delta^2=-50~\text{dBm}$    \\ \hline
		Rician factor & 5 dBw \\ \hline
	\end{tabular}
\end{table}

\section{Numerical Results}
In this section, numerical results are provided to assess the improvements made by the proposed algorithms. An IRS-SWIPT femtocell network is considered where single-antenna users are distributed uniformly in a square area whose side is 5m. The system parameters are summarized in Table 1. The following equation is used for the path loss model:
\begin{equation}
L(d)=C_0\left(\frac{d}{D_0}\right)^{-\alpha},
\end{equation}
where $C_0$ denotes the path loss at the reference distance $D_0 =1$m, which is set to be $-30$ dBw, $\alpha$ indicates path loss exponent considered to be 2, and $d$ indicates the length of the link. For all the included channel links, the Rician fading model with the dominant line-of-sight (LOS) signal is considered as follows:  
\begin{align}
&\mathbf{h}_{i,k}=\sqrt{\frac{K_r}{1+K_r}}\mathbf{h}_{i,k}^{\text{LOS}}+\sqrt{\frac{1}{1+K_r}}\mathbf{h}_{i,k}^{\text{NLOS}},\: i\in \{b,r\},\label{key10}\\
&\mathbf{H}_{b,r}=\sqrt{\frac{K_r}{1+K_r}}\mathbf{H}_{b,r}^{\text{LOS}}+\sqrt{\frac{1}{1+K_r}}\mathbf{H}_{b,r}^{\text{NLOS}},\label{key11}
\end{align}
where $\mathbf{h}_{i,k}^{\text{LOS}}$ and $\mathbf{H}_{b,r}^{\text{LOS}}$ are the LOS components, and the Rayleigh fading component is expressed by $\mathbf{h}_{i,k}^{\text{NLOS}}$ and $\mathbf{H}_{b,r}^{\text{NLOS}}$ whose elements follow a Gaussian random variable with zero mean and covariance one. Furthermore, $K_r$ is a Rician factor, which is assumed to be the same for all models. Particularly, a uniform linear antenna array (ULA) model is adopted for the LOS component, i.e., $\mathbf{h}_{i,k}^{\text{LOS}}=[1, e^{j\theta_k},..., e^{j(M-1)\theta_k}]^T$ with $\theta_k=-\frac{2\pi x\sin(\phi_k)}{\lambda}$, where $\lambda$ denotes the the carrier wavelength and $x=\lambda/2$ is the distance between two consecutive antennas. Similarly, $\mathbf{H}_{b,r}^{\text{LOS}}$ can be modeled by the ULA. Finally, the weight $\lambda$ in the objective of P5 was set to one.

We investigate the improvement made by IRS presence compared to when IRS is absent. In this regards, the optimal PS-SWIPT parameters derived by \cite{12} are computed and labeled as optimal solution without IRS. Similarly, ZF without IRS is also borrowed from \cite{12}. As for MRT without IRS, only P8 is solved once. In all three cases, channel gains $\bh_k$ are set equal to $\bh_{b,k}$, which is the direct channel between the BS and users. These results are compared against our three proposed algorithms when the IRS exists.

\begin{figure}[t]
	\centering
	\includegraphics[width=0.8\textwidth]{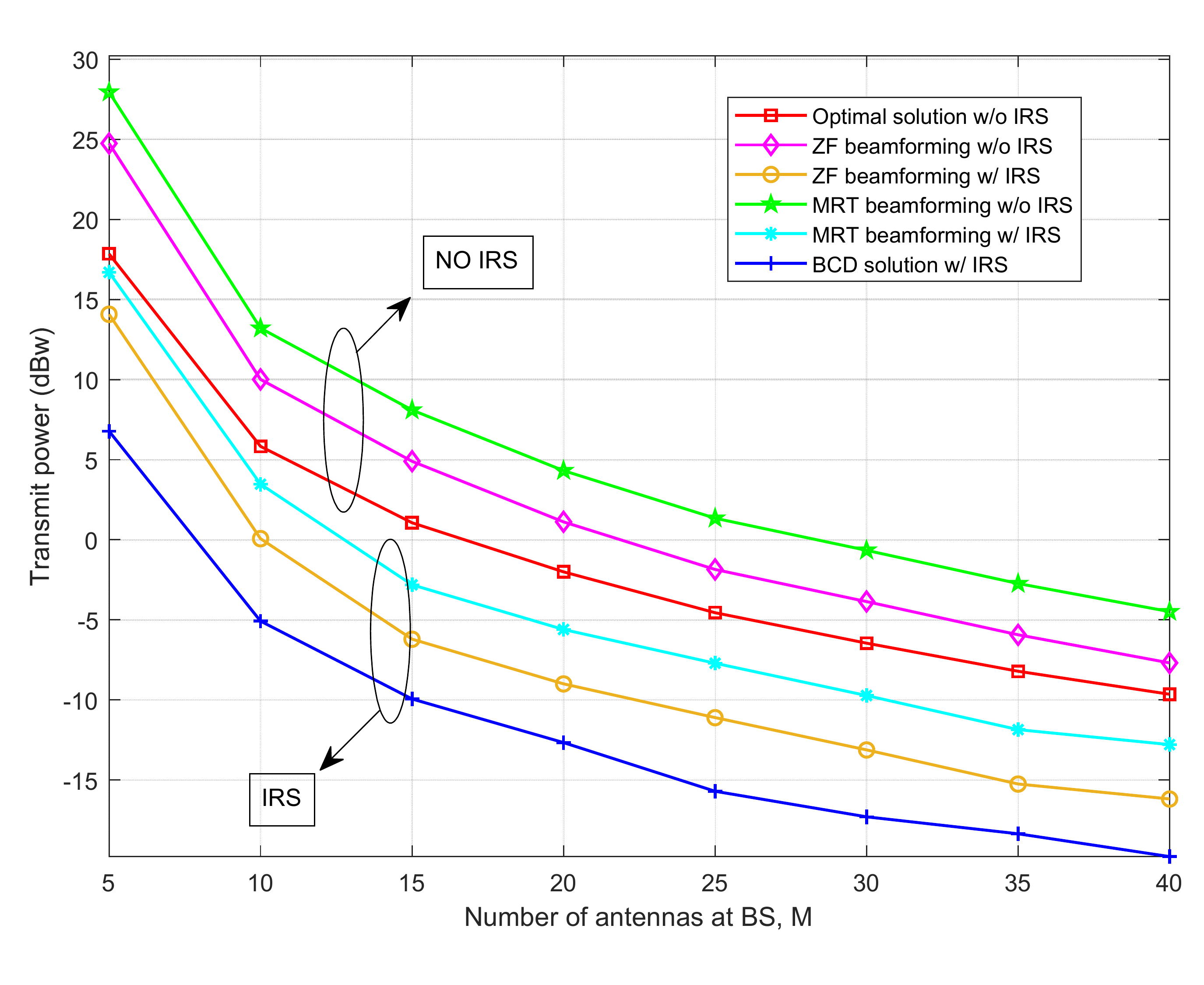}
	\caption{{\small Transmission power versus $M$ with fixed $e_m=-10$ dBm and $\gamma_k=10$ dBw, $N=50$.}} \label{fig:versus_m}
\end{figure}

First, we look at the minimum required transmit power at the BS against the number of BS antennas $M$, while $\gamma_k=10$ dBw and $e_k=-10$ dBm are fixed. As shown in Fig \ref{fig:versus_m}, it can be observed that the transmit power at the BS decreases when the number of transmit antennas at the BS grows in all our proposed algorithms. This fact reveals that exploitation of large or even massive antennas can be advantageous for MISO IRS-aided SWIPT systems in order to reduce the transmit power. As seen, the proposed BCD Algorithm 1 performs better than the globally optimal design without IRS. It can also be observed that even the sub-optimal MRT- and ZF-based beamformers with IRS can enhance the system performance compared to that of the non-IRS case.

\begin{figure}[t]
	\centering
	\includegraphics[width=0.8\textwidth]{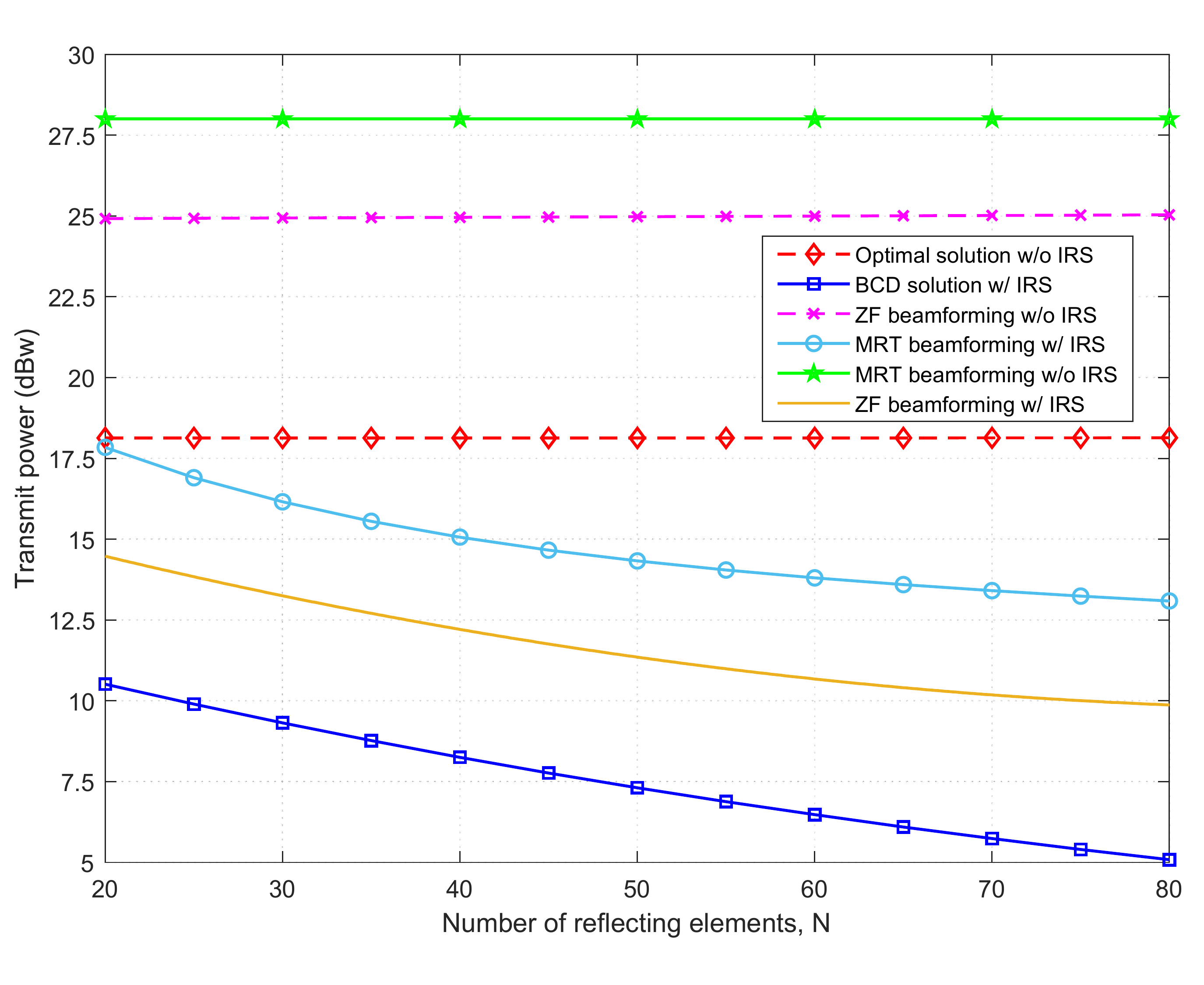}
	\caption{{\small BS transmit power versus the number of passive elements at the IRS, $N$, with fixed $e_m=-$10 dBm and $\gamma_k=10$ dBw.}} \label{fig:versus_n}
\end{figure}

\begin{figure}[h!]
	\centering
	\includegraphics[width=0.8\textwidth]{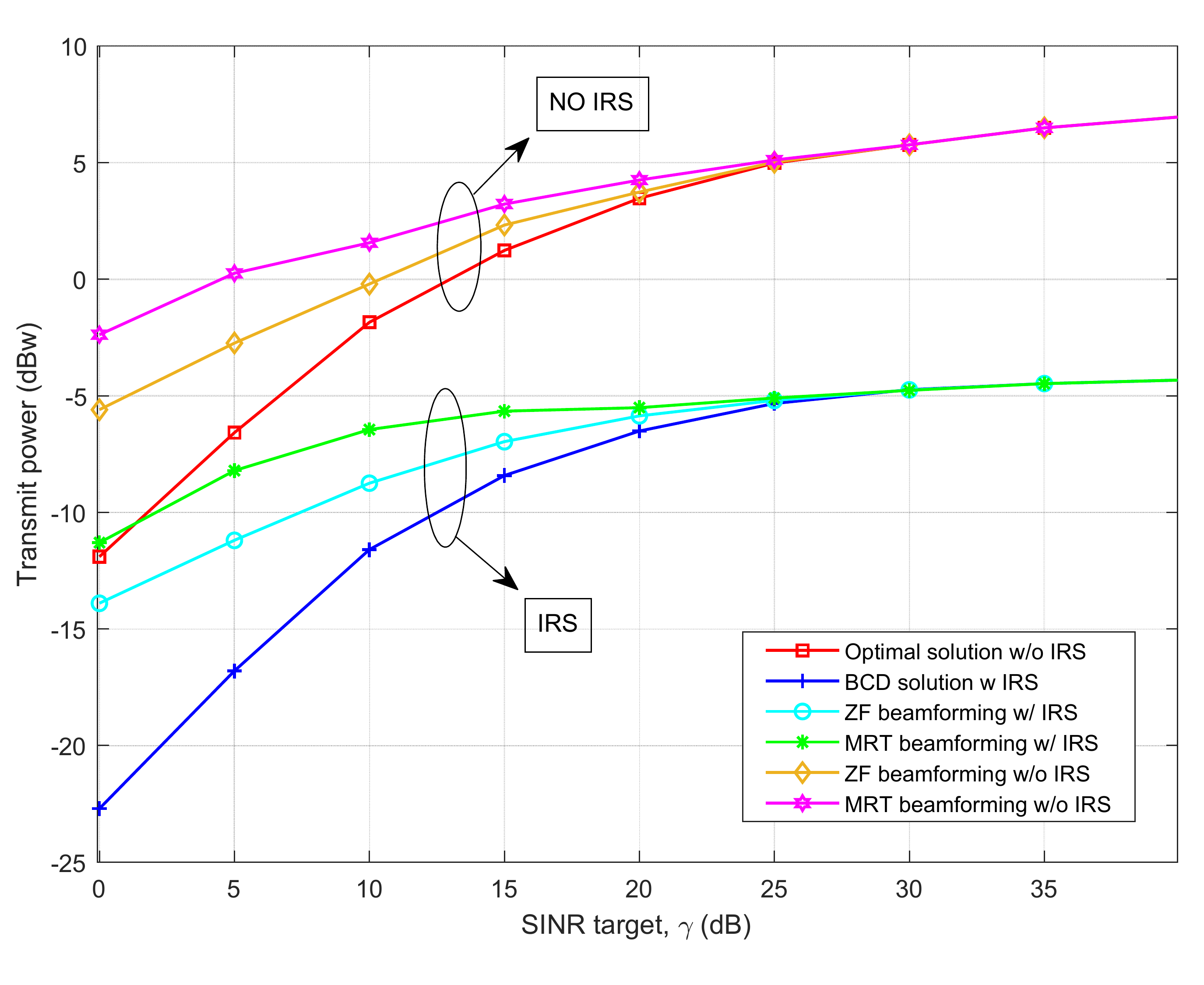}
	\caption{{\small  Transmit power versus SINR target $\gamma$, with fixed $e_m=-$20 dBm and $N=50$.}}\label{fig:versus_gamma}
\end{figure}

Next, we plot minimum transmit power versus the number of IRS passive components $N$ for a fixed $\gamma_k=10$ dBw and $e_k=-10$ dBm in Fig. \ref{fig:versus_n}. Similar to Fig. \ref{fig:versus_m}, the BS transmit power decreases by increasing the number of units at the IRS. In fact, deploying IRS with more passive elements can further amplify the signal power and therefore reduce the required transmit power at the BS. Nevertheless, in the non-IRS case, the transmit power does not change and stays the same in optimal, MRT, and ZF sub-optimal solutions as expected. Indeed, provided enough IRS elements are employed, a sub-optimal MRT/ZF beamformer with IRS can outperform the globally optimal design without IRS. This is a testament to the effectiveness of the  IRS-aided communications. Finally, Fig. \ref{fig:versus_gamma} shows the performance comparison between the IRS and non-IRS cases with BCD, ZF, and MRC based sub-optimal solutions, where $e_k=-20$ dBm is fixed, and the BS is equipped with $M=4$ antennas. When IRS is present, it has $N=50$ elements. It is observed that the transmit power is an increasing function of minimum required SINR $\gamma_k$. In all cases, the IRS-aided SWIPT system performs more satisfactory than non-IRS cases. Besides, the BCD solution obtains a very smaller transmission power in the lower SINR region than the ZF and MRC based sub-optimal solutions. However, as SINR increases, the gap between the two disappears. Low number of BS antennas in this scenario suggests that IRS can be beneficial for device to device (D2D) communications as well. Indeed, a preinstalled IRS platform can help a transmitter with few antennas to reduce its transmit power significantly as if it is using a massive array.

\section{Conclusions}
An optimization problem to jointly select the BS beamforming vectors, IRS phase shifts, and receivers PS ratios for a multiuser MISO IRS-aided SWIPT-based cell was formulated subject to minimum rate and harvested energy QoS constraints at the receivers. A block coordinate descent algorithm was proposed with the two desirable properties of decreasing the objective at every iteration and guaranteed convergence. Furthermore, two sub-optimal algorithms with a considerably lower complexity were developed by utilizing MRT- and ZF-based beamforming. Numerical results revealed the significant performance improvement of the proposed approaches compared to the global optimum of a similar SWIPT system with no IRS.

\section*{Appendix A: Proof of Proposition 1}
We consider $\{\mathbf{w}_k^{(i+1)},\rho_k^{(i+1)}\}_{k=1}^K$ as the optimal solution of P2 given $\mathbf{\Theta}^{(i)}$. Furthermore let $\mathbf{\Theta}^{(i+1)}$ be the optimal solution of P5, given $\{\mathbf{w}_k^{(i+1)},\rho_k^{(i+1)}\}_{k=1}^K$. If we define the objective function of P1 as
\[
f\left(\bTet,\{\bw_k,\rho_k\}_{k=1}^K\right)=\sum_{k=1}^K~\|\bw_k\|^2.
\]
Then, we will have
\begin{eqnarray*}
	f\left(\bTet^{(i+1)},\{\bw_k^{(i+1)},\rho_k^{(i+1)}\}_{k=1}^K\right)\\ =  f\left(\bTet^{(i)},\{\bw_k^{(i+1)},\rho_k^{(i+1)}\}_{k=1}^K\right) \\  \leq  f\left(\bTet^{(i)},\{\bw_k^{(i)},\rho_k^{(i)}\}_{k=1}^K\right) 
\end{eqnarray*}
which comes from the fact that for given $\mathbf{\Theta}^{(i)}$, the solutions $\{\mathbf{w}_k^{(i+1)},\rho_k^{(i+1)}\}_{k=1}^K$ are optimal. As for convergence, we have a sequence of non-increasing objective values, which are bounded below by zero. Note that the objective function can not become negative. Hence the sequence of objective values are guaranteed to converge.

\bibliographystyle{elsarticle-num}      % basic style, author-year citations
\bibliography{refrence}   % name your BibTeX data base

\begin{thebibliography}{10}
\expandafter\ifx\csname url\endcsname\relax
  \def\url#1{\texttt{#1}}\fi
\expandafter\ifx\csname urlprefix\endcsname\relax\def\urlprefix{URL }\fi
\expandafter\ifx\csname href\endcsname\relax
  \def\href#1#2{#2} \def\path#1{#1}\fi

\bibitem{survey1}
T.~Huang, W.~Yang, J.~Wu, J.~Ma, X.~Zhang, D.~Zhang, {A Survey on Green 6G
  Network: Architecture and Technologies}, IEEE Access 7~(9) (2019)
  175758--175768.

\bibitem{survey2}
X.~Yuan, Y.-J. Zhang, Y.~Shi, W.~Yan, H.~Liu,
  {Reconfigurable-Intelligent-Surface Empowered 6G Wireless Communications:
  Challenges and Opportunities}, arXiv preprints 2001.00364 (Jan. 2020).

\bibitem{survey3}
S.~{Zhang}, Q.~{Wu}, S.~{Xu}, G.~Y. {Li}, {Fundamental Green Tradeoffs:
  Progresses, Challenges, and Impacts on 5G Networks}, IEEE Communications
  Surveys and Tutorials 19~(1) (2017) 33--56.

\bibitem{survey4}
Q.~{Wu}, G.~Y. {Li}, W.~{Chen}, D.~W.~K. {Ng}, R.~{Schober}, {An Overview of
  Sustainable Green 5G Networks}, IEEE Wireless Communications 24~(4) (2017)
  72--80.

\bibitem{2}
E.~{Basar}, M.~{Di Renzo}, J.~{De Rosny}, M.~{Debbah}, M.~{Alouini},
  R.~{Zhang}, {Wireless Communications Through Reconfigurable Intelligent
  Surfaces}, IEEE Access 7 (2019) 116753--116773.

\bibitem{1}
Q.~{Wu}, R.~{Zhang}, {Towards Smart and Reconfigurable Environment: Intelligent
  Reflecting Surface Aided Wireless Network}, IEEE Communications Magazine
  58~(1) (2020) 106--112.

\bibitem{3}
H.~Zhang, B.~Di, L.~Song, Z.~Han, {Reconfigurable Intelligent Surfaces Assisted
  Communications With Limited Phase Shifts: How Many Phase Shifts Are Enough?},
  IEEE Transactions on Vehicular Technology 69~(4) (2020) 4498–4502.

\bibitem{5}
C.~{Huang}, A.~{Zappone}, M.~{Debbah}, C.~{Yuen}, Achievable rate maximization
  by passive intelligent mirrors, in: {IEEE International Conference on
  Acoustics, Speech and Signal Processing (ICASSP)}, Calgary, AB, Canada, 2018,
  pp. 3714--3718.

\bibitem{7}
C.~{Huang}, A.~{Zappone}, G.~C. {Alexandropoulos}, M.~{Debbah}, C.~{Yuen},
  {Reconfigurable Intelligent Surfaces for Energy Efficiency in Wireless
  Communication}, IEEE Transactions on Wireless Communications 18~(8) (2019)
  4157--4170.

\bibitem{8}
X.~Yu, D.~Xu, R.~Schober, {MISO Wireless Communication Systems via Intelligent
  Reflecting Surfaces : (Invited Paper)}, in: 2019 IEEE/CIC International
  Conference on Communications in China (ICCC), Changchun, China, 2019, pp.
  735--740.

\bibitem{12}
Q.~{Shi}, L.~{Liu}, W.~{Xu}, R.~{Zhang}, {Joint Transmit Beamforming and
  Receive Power Splitting for MISO SWIPT Systems}, IEEE Transactions on
  Wireless Communications 13~(6) (2014) 3269--3280.

\bibitem{11}
H.~{Lee}, S.~{Lee}, K.~{Lee}, H.~{Kong}, I.~{Lee}, {Optimal Beamforming Designs
  for Wireless Information and Power Transfer in MISO Interference Channels},
  IEEE Transactions on Wireless Communications 14~(9) (2015) 4810--4821.

\bibitem{10}
J.~{Xu}, L.~{Liu}, R.~{Zhang}, {Multiuser MISO Beamforming for Simultaneous
  Wireless Information and Power Transfer}, IEEE Transactions on Signal
  Processing 62~(18) (2014) 4798--4810.

\bibitem{13}
M.~J. Emadi, H.~Masoumi, {Performance Analysis of cooperative SWIPT System:
  Intelligent Reflecting Surface versus Decode-and-Forward}, AUT Journal of
  Modeling and Simulation (2019).

\bibitem{14}
Y.~Tang, G.~Ma, H.~Xie, J.~Xu, X.~Han, {Joint Transmit and Reflective
  Beamforming Design for IRS-Assisted Multiuser MISO SWIPT Systems}, arXiv
  preprints 1910.07156 (2019).

\bibitem{15}
Q.~{Wu}, R.~{Zhang}, {Weighted Sum Power Maximization for Intelligent
  Reflecting Surface Aided SWIPT}, IEEE Wireless Communications Letters 9~(5)
  (2020) 586--590.

\bibitem{16}
Q.~{Wu}, R.~{Zhang}, {Joint Active and Passive Beamforming Optimization for
  Intelligent Reflecting Surface Assisted SWIPT under QoS Constraints}, arXiv
  preprints 1910.06220v2 (Jul. 2020).

\bibitem{original_beamforming}
M.~Bengtsson, B.~Ottersten, {Optimal and suboptimal transmit beamforming}, in:
  L.~C. Godara (Ed.), Handbook of Antennas in Wireless Communications, 2002,
  Ch.~18.

\bibitem{np-hard}
Y.~{Liu}, Y.~{Dai}, Z.~Q. {Luo}, {Coordinated Beamforming for MISO Interference
  Channel: Complexity Analysis and Efficient Algorithms}, IEEE Transactions on
  Signal Processing 59~(3) (2011) 1142--1157.

\bibitem{17}
S.~Boyd, L.~Vandenberghe, {Convex Optimization}, Cambridge University Press,
  2004.

\bibitem{18}
M.~Grant, S.~Boyd, {{CVX}: Matlab Software for Disciplined Convex Programming},
  \url{http://cvxr.com/cvx} (2014).

\bibitem{19}
S.~{Timotheou}, I.~{Krikidis}, G.~{Zheng}, B.~{Ottersten}, {Beamforming for
  MISO Interference Channels with QoS and RF Energy Transfer}, IEEE
  Transactions on Wireless Communications 13~(5) (2014) 2646--2658.

\end{thebibliography}





\end{document}